\documentclass[12pt]{article}

\usepackage[height=25cm,a4paper,hmargin={2cm,2cm}]{geometry}

\usepackage{xcolor}

\usepackage{url}

\usepackage{amsfonts}
\usepackage{amssymb}
\usepackage{amsmath}

\usepackage{mathtools}
\DeclarePairedDelimiter\bra{\langle}{\rvert}
\DeclarePairedDelimiter\ket{\lvert}{\rangle}
\DeclarePairedDelimiterX\braket[2]{\langle}{\rangle}{#1 \delimsize\vert #2}

 \usepackage{xcolor}

\usepackage { boxedminipage }

\usepackage{setspace}
\def\C{\mathbb C}
\def\Curl{\mathrm{curl}}
\def\TE{\mathrm{T}_{\varepsilon,\mu}}

\def\lambdas{\lambda_s}
\def\omegas{\omega_s}
\def\KOmega{\mathcal{D}}
\def\KU{{\mathcal{U}}}

\definecolor{colorgui}{RGB}{0, 0, 0}
\newcommand{\gui}[1]{\textcolor{colorgui}{#1}}
\newcommand{\cg}[1]{\textcolor{black}{#1}}

\definecolor{colorrewier1}{RGB}{0,0,0}
\definecolor{colorrewier2}{RGB}{0,0,0}
\definecolor{colorme}{RGB}{0,0,0}
\def\crevi{\color{colorrewier1}}
\def\crevii{\color{colorrewier2}}

\usepackage{authblk}
\title{\textbf{Physically Agnostic Quasinormal Mode Expansion in Time Dispersive Structures:\\ from Mechanical Vibrations \\to Nanophotonic Resonances}}
\author[1]{Andr\'e Nicolet}
\author[1]{Guillaume Dem\'esy}
\author[1]{Fr\'ed\'eric Zolla}
\author[2]{Carmen Campos}
\author[3]{Jose E. Roman}
\author[4]{Christophe Geuzaine}
\affil[1]{Aix Marseille Univ, CNRS, Centrale Med, Institut Fresnel, Marseille, France}
\affil[2]{Universitat de Val\`encia, D. Did\'actica de la Matem\'atica,
Avda. Tarongers, 4,   \newline Val\`encia, E-46022, 
Spain}
\affil[3]{Universitat Polit\`ecnica de Val\`encia, D. Sistemes Inform\`atics i Computaci\'o,
            Cam\'{\i} de Vera, s/n, 
           Val\`encia,
          E-46022, 
	Spain}
\affil[4]{University of Li\`ege, Dept. of Electrical Engineering and Computer Science,\newline
            All\'ee de la D\'ecouverte~10, 
            Li\`ege,
        B-4000, 
	Belgium}
\date{}                     
\begin{document}
\let\WriteBookmarks\relax
\def\floatpagepagefraction{1}
\def\textpagefraction{.001}





\maketitle

\begin{abstract}
Resonances, also known as quasi normal modes (QNM) in the non-Hermitian case, play an ubiquitous role in all domains of physics ruled by wave phenomena, notably in continuum mechanics, \gui{acoustics}, electrodynamics, and quantum theory. In this paper, we present a QNM expansion for dispersive systems, recently applied to photonics but based on sixty year old techniques in mechanics. The resulting numerical algorithm appears to be physically agnostic, that is independent of the considered physical problem and can therefore be implemented as a mere toolbox in a nonlinear eigenvalue computation library.
\end{abstract}

\section{Introduction}
\subsection{Resonances and quasi normal modes (QNM) in physics}
Quasi normal modes (QNM) are a powerful tool to analyze wave problems in physics. Thanks to the seminal work of Joseph Fourier in his ``Th{\'e}orie analytique de la chaleur'' \cite{Fourier1822theorie} in 1822, it has been natural in the 20th century to consider problems from both the direct position/time point of view as well as from the \gui{momentum/frequency} dual point of view using the Fourier transform that is provided by the decomposition of a function in terms of plane waves, \textit{i.e.} modes (solution without sources) of wave equations in free space. 
For a particular problem associated to a given structure described by its geometry, boundary conditions, and media properties, the response to a solicitation with a given frequency is strongly correlated to the \gui{intrinsic} properties of the system and it appears that the strongest responses are related to the \textbf{resonances} of the structure, \textit{i.e.}\ solutions of the wave equation without sources, no more in free space but associated to the particular problem. It appears that these solutions are eigenmodes of the corresponding particular operator and the set of these eigenmodes is a particularly well suited basis to develop other solutions with a given source. Determining these eigenmodes could therefore be extremely useful for both the physical understanding and for practical computations. It is also expected that small subsets of these modes could contain enough information to deal with some problems and constitute an efficient reduced model for them.

A dramatic and popular example of the importance of resonances is the collapse of the
Tacoma Narrows Bridge that is nevertheless due to much more complex phenomena~\cite{Billah1991ResonanceTN}. Much more recent cases are the Gateshead Millennium Bridge nicknamed the ``Wobbly Bridge'' after pedestrians experienced an alarming swaying motion on its opening day and the Volga Bridge in Volgograd~\cite{Brun_2013}. 
New methods have been designed to prevent these catastrophic vibrational damages to occur because of resonances. \gui{Conversely, resonances can be used to} \cg{design and study novel} metamaterials and photonic/phononic crystals~\cite{movchan2016metamaterial}.

Another example of modes are the propagating modes in waveguides such as optical fibres. In the early 2000s\cg{, microstructured} optical fibres \gui{have} appeared. The initial idea was to use the bandgap of a photonic crystal fibre, but it became soon clear that a limited number of periodic holes in the cladding \cg{were} enough to obtain good guiding properties~\cite{pcf_foundations}. A basic model is to consider low refractive index holes in a higher refractive index, large enough to be considered as unbounded. In this case, there are no genuine \gui{propagative} modes but rather leaky modes associated to complex eigenvalues \gui{(\emph{i.e.} propagation constants)}. These modes \cg{do} suffer \cg{from} losses, but \cg{which are} small enough to preserve excellent guiding properties. More generally, the materials used in photonics are represented by a complex dielectric permeability where the imaginary part \cg{corresponds} to losses. All the classical optical materials at optical frequencies are dispersive,  \textit{i.e.}\ frequency dependent, and are therefore dissipative in accordance to the Kramers-Kronig relations~\cite{lucarini2005kramers} induced by the principle of causality.

\cg{In view of the above, we understand that considering unbounded problems and realistic models for media leads to frequency-dependent non-Hermitian operators, with the non-hermiticity due to both the losses in the materials and the leakage at infinity}. In fact, a very common way to deal with unbounded wave problems, in order to simulate the outgoing wave conditions, is to introduce absorbing wave conditions at finite distance that involve some complex coefficients. The imaginary part of these coefficients is taking into account the outgoing power never coming back. Such techniques were already used in the early 70s in quantum mechanics,  \textit{i.e.}\ the dilatation-analytic method, to determine atomic and molecular resonances~\cite{BaslevCombes} and are really similar to the Perfectly Matched Layers (PML) introduced in computational electrodynamics ~\cite{pml_berenger,YaYanLuPML1,goursaudPML,pmlCompel}.

The idea of computing modes for resonators and waveguides can then be extended to arbitrary structures, even with small quality factors: this is the \textbf{quasi normal mode} analysis. The name ``quasi normal mode'' emphasizes the fact that the considered operator is non-Hermitian hence that its eigenvalues are complex and the orthogonal relations associated to Hermitian operators are not available. Note \cg{that} non-Hermitian problems in physics  are currently receiving a lot of attention \cite{ashida2020non}.

\subsection{The dispersive modal expansion}\label{part:Keldysh}

As an example of physical problem to present the concepts of dispersive modal expansion,  we consider an electromagnetic system ruled by a set of partial differential equations obtained by combining Maxwell's equations with constitutive relations for media (and giving boundary/transmission/outgoing wave conditions), since, in the last decade, this modal expansion formalism has received a lot of attention in photonics  \cite{bai2013efficient,sauvan2013theory,vial2014quasimodal,muljarov2016resonant,yan2018rigorous,BinkowskiQNM2020,
BinkowskiRiesz2020,Lalanne:19}. 
Developing a solution in terms of modes is a well known technique~\cite{hanson2013operator}. 
As we have just seen above, practical use requires to take into account realistic time dispersive media and this leads to non-Hermitian nonlinear eigenvalue problems~\cite{spence2005photonic,engstrom2017rational}. 
Nowadays, efficient numerical algorithms are available for the numerical computations of such problems ~\cite{Hernandez:2005:SSF,voss2013nonlinear,guttel2017nonlinear}. The use of these nonlinear eigenmodes in modal expansions is still an active research topic.
As for the modal expansion, new formulations had to be \gui{derived} \cite{doost2015resonant, doost2016resonant,muljarov2016resonant,doost2017resonant,yan2018rigorous,BinkowskiQNM2020,
BinkowskiRiesz2020}. Most of the work has been performed in the framework of Maxwell's equations formalism with some physical interpretation in mind, using for instance conjugated and unconjugated forms of the reciprocity theorem as it is proposed in the Chap. 31 ``Modal methods for Maxwell's equations'' of \cite{snyder1983optical} and several formulas have been proposed by various research teams.

In order to find a general formula based on established theorems, the authors of the present paper have adopted a more abstract approach, \textit{i.e.}\ independent of the physical context and therefore applicable to a large range of physical models. In \cite{zolla2018photonics}, we have proposed a straightforward approach to the modal expansion in the dispersive case for photonics based on the Keldy\v{s} theorem ~\cite{keldysh1951eigenvalues,keldysh1971completeness,kozlov1999differential,beyn2012integral,van2016nonlinear,unger2013convergence} that provides modal expansions for very general systems (both permittivity and permeability may be dispersive, anisotropic, and even non reciprocal). 
In our approach, the central concept is the set of the eigentriplets associated to the nonlinear eigenproblem and no normalization is involved (see sections \ref{eigentriplets} and \ref{pencils}  below for these notions and the related notations). In the simplest cases, the expansion is similar to recent results for dispersive permittivity ~\cite{leung1994completeness,ge2014quasinormal,perrin2016eigen,muljarov2016resonant,doost2016resonant,lalanne2018light,yan2018rigorous}. 

Given a particular system, and here the physical nature of the system is forgotten in favour of its mathematical formalisation, we consider the corresponding holomorphic operators $T(z)$, that is a set of non-Hermitian operators parametrized by a complex parameter $z$~\cite{kozlov1999differential} 
and the associated eigenvalue problem:
given a holomorphic operator $T(z)$, we define its \textit{eigentriplets} as ordered sets $(\lambda ,\bra{\mathbf{u_l}}, \ket{\mathbf{u_r}})$ of an eigenvalue and its associated \textit{left} and \textit{right} eigenvectors satisfying $\bra{\mathbf{u_l}} T(\lambda)=0$ and $T(\lambda) \ket{\mathbf{u_r}} =0$. An eigenvalue is simple if $\bra{\mathbf{u_l}} T'(\lambda) \ket{\mathbf{u_r}}  \neq 0$ where $T'(z)$ is the complex derivative of $T(z)$ and, in the sequel, we will consider that all the eigenvalues are simple or at least semi-simple 
and that the operators are diagonalizable. In practice, finding the couples $(\lambda ,\bra{\mathbf{u_l}})$ satisfying $\bra{\mathbf{u_l}} T(\lambda)=0$ amounts to looking for \textit{right} eigenvectors $\ket{\mathbf{u_l}}$ such that $T^*(\lambda) \ket{\mathbf{u_l}} =0$ where $T^*$ is the adjoint operator associated with $T$ and, in the sequel, $\bra{\mathbf{u_l}} \ne \bra{\mathbf{u_r}}$ due to the non self-adjointness of the operators at stake.

In this case, the modal expansion is provided by the Keldy\v{s} theorem \cite{keldysh1951eigenvalues,keldysh1971completeness}. This theorem from the 1950s has recently found very useful applications in the nonlinear eigenvalues numerical computation algorithm.  Indeed, for the sake of simplicity, we give its version for matrices with simple eigenvalues as stated by Van Barel and Kravanja \cite{van2016nonlinear} in their presentation of Beyn's algorithm \cite{beyn2012integral} (\gui{a contour integral} method for solving nonlinear eigenvalue problems):

\textbf{Theorem} (Keldy\v{s}):
Given a domain $\KOmega\subset\C$ and an integer $m \geq 1$, let $\mathcal{C} \subset \KOmega$ be a compact subset, let $T$ be a matrix-valued function $T: \KOmega \longrightarrow \C^{m \times m}$ analytic in $\KOmega$
and let $n(\mathcal{C})$ denote the number of eigenvalues of $T$ in $\mathcal{C}$.
Let $ \lambda_k$ for $k = 1, \ldots, n(\mathcal{C})$  denote these eigenvalues and suppose that all of them are simple. Let $\ket{\mathbf{u_l}_{k}}$ and $\bra{\mathbf{u_r}_k}$ for $k = 1, \ldots, n(\mathcal{C})$ denote their left and right eigenvectors, such that
\[
T(\lambda_k) \ket{\mathbf{u_r}_{k}} =0 ,  \; \; \; \; \; \; \; \; \;    \bra{\mathbf{u_l}_k} T(\lambda_k)=0.
\]
Then there is a neighborhood $\KU$ of $\mathcal{C}$ in  $\KOmega$ and a matrix-valued analytic function $R: \KU \longrightarrow \C^{m \times m}$ such that the resolvent $T(z)^{-1}$ can be written as
\begin{equation}
T(z)^{-1}=  \sum_{k=1}^{n(\mathcal{C})} \frac{1}{(z-\lambda_k)} \frac{\ket{\mathbf{u_r}_{k}} \,   \bra{\mathbf{u_l}_k}}{\braket{\mathbf{u_l}_k }{T'(\lambda_k) \mathbf{u_r}_k } }+ R(z) \label{keldysh}
\end{equation} 
for all $z \in \KU \setminus \{\lambda_1 ,\ldots, \lambda_{n(\mathcal{C})} \}$.
If $T$ is a matrix-valued strictly proper rational function, the analytic function $R$ is equal to zero~\cite{van2016nonlinear}.

Considering the similar results for holomorphic operator functions does not cause additional difficulties~\cite{kozlov1999differential}. 

Our conclusion is that the best approach is \gui{to} use very general mathematical tools, such as the Keldy\v{s} theorem, largely independent of the physical context.
Moreover, for all practical purposes, the modal expansion will be performed at the discrete level. Therefore, in the sequel, we will consider matrix problems arising from the discretisation of partial differential operators using appropriate numerical methods. \gui{Note that this framework naturally applies when the discretisation is independent of the frequency such as in Finite Element Method and Finite Difference Method, but unlike methods using Green function or Fourier-based basis. Nevertheless some authors are efficiently using Boundary Element Method in a similar context \cite{unger2018novel,BinkowskiQNM2020}, which allows to leave the frequency (eigenvalue) outside of the large matrices (discrete operators)}. Considering this step granted, we can apply sixty-year-old techniques introduced by Lancaster for the study of mechanical vibrations~\cite{lancaster1960inversion,lancaster1966lambda}. Despite the old-fashioned terminology \textit{e.g.}\ ``lambda matrices'' for the matrices $T(z)$ with coefficients that are polynomials of a variable $z$ and ``latent root'' $\lambda$ and ``latent vector'' $\mathbf{v}$ such that $T(z) \mathbf{v} = \mathbf{0}$ that we will just name (nonlinear) eigenvalue and eigenvector, the theory is directly applicable to the quasi normal mode expansion with a slight modification to take into account the fact that the coefficients are in fact rational functions of $z$.

\section{Non-Hermitian eigenproblems}

\subsection{The eigentriplets\label{eigentriplets}}

Instead of considering the general case of linear operators that are linear maps between Banach spaces, we will consider the finite dimensional case since our purpose is numerical modeling with matrices as stated above.
Given a linear operator represented in a particular basis by the matrix $A$, a \textbf{(right) eigenpair} $(\lambda, \mathbf{v})$ is an ordered pair of a complex number  $\lambda$ and a complex vector (an element of the domain space) $\mathbf{v} \neq \mathbf{0}$ such that $$A \mathbf{v} = \lambda \mathbf{v}.$$
Only the direction of $\mathbf{v}$ is relevant since for any $\alpha \in \C \neq 0$, $(\lambda, \alpha \mathbf{v})$ is an equivalent eigenpair. An eigenvector $\mathbf{v}$ can be normalized but this has no particular meaning, and a particular eigenvector is a mere description of a 1-dimensional eigenspace. The eigenpair is an intrinsic characteristic of the linear operator represented by the matrix $A$ in a particular basis {\textit{i.e.}} the eigenpair does not depend on the basis choice except, of course, for the values of the components of $\mathbf{v}$ given in the current basis.
A \textbf{left eigenpair} $(\lambda, \mathbf{u}^H)$ can also be defined such that $$ \mathbf{u}^H A = \lambda \mathbf{u}^H,$$
where the superscript $H$ indicates the difference of nature between the left eigenvectors and the right eigenvectors. In the matrix algebra context, $\mathbf{v}$ is a column vector of coefficients while $\mathbf{u}^H$ is a row vector of coefficients where the superscript $H$ denotes the Hermitian conjugate (complex conjugate of the transposed matrix)  \textit{i.e.}\ the operation used to turn a column vector $\mathbf{v}$ into a row vector $\mathbf{v}^H$ of its complex conjugate components. Behind the shallow notational aspect, $\mathbf{v}$ is a contravariant vector that may also be denoted as a ket $\ket{\mathbf{v}}$ in Dirac bra-ket notation while $\mathbf{u}^H$ is a covariant vector that can be denoted as a bra $\bra{\mathbf{u}}$. The duality product can be written as a matrix product $\mathbf{u}^H \mathbf{v}$ as well as a bra-ket $\braket{\mathbf{u}}{\mathbf{v}}$.

The left eigenvector is in fact the classical right eigenvector of the adjoint operator $A^*$ defined by $\braket{\mathbf{u}}{A\mathbf{v}}=\braket{A^*\mathbf{u}}{\mathbf{v}}$. In the matrix case, computing $(\mathbf{u}^H A)^H $ gives $$ A^H \mathbf{u} = \overline{\lambda} \mathbf{u}.$$

Consider now the case where the matrix is not defective \textit{i.e.} reducible to a diagonal matrix. The necessary and sufficient condition is that for a matrix of dimension $n$, the number of linearly independent eigenvectors is equal to $n$. It means that for a multiple eigenvalue, its geometric multiplicity is always equal to its algebraic multiplicity and this is always the case when all the eigenvalues are simple. In this case, the column right eigenvectors can be gathered in a square invertible matrix $R$ such that $AR =  R\Lambda$ and therefore
 $R^{-1}AR = \Lambda$ where $\Lambda$ is the diagonal matrix built with the eigenvalues ordered accordingly to the order chosen to place the eigenvectors as columns of $R$. Note that any eigenvector can be multiplied by an arbitrary nonzero complex number (as already noted, it is indeed more appropriate to speak about a one dimensional eigensubspace) and the matrix $R$ can be multiplied by any nonsingular diagonal matrix $D$ to get another matrix that also diagonalizes $A$.

A similar process can be performed placing the left eigenvectors: building $L^H$ with the row left eigenvectors, keeping of course the same order for the eigenvalues, to get $L^H A = \Lambda L^H$ and $L^H A L^{-H}= \Lambda$ where $L^{-H}$ is the inverse of $L^H$. Therefore we have $L^H A L^{-H}= \Lambda = R^{-1}AR = D R^{-1} A R D^{-1} $. This gives the equality 
$$L^H R = D.$$
The right and left vectors form two \textbf{biorthogonal} sets. Columns of $R$ form a basis for the contravariant vectors and rows of $L^H$ form its \textbf{dual basis} for the covariant vectors. There is no natural normalization other than choosing $D=I$ but this is a constraint on the $L^H$ and $R$ relation and leaves the freedom of choice for the amplitudes of the eigenvectors. The fact that $R$ and $L^H$ are dual bases for vectors and covectors indicates that they are in one-one correspondence and there is no room for other linearly independent left eigenvectors possibly associated to another eigenvalue. With the $D=I$ choice, $L^H=R^{-1}$ and, from $AR=R\Lambda$, we have $(AR)^{-1}=(R\Lambda)^{-1}$, then $R ^{-1} A ^{-1} = \Lambda ^{-1} R ^{-1}$, $\Lambda R ^{-1} = R ^{-1} A$ and, finally, $\Lambda L^H = L^H A$ and $L^H$ is indeed associated to the same eigenvalues than $R$.  

Therefore we can define \textbf{eigentriplets} as the ordered sets $(\lambda ,\mathbf{u}^H, \mathbf{v})$ of an eigenvalue and its associated left and right eigenvectors.  

A common particular case are the Hermitian matrices $A=A^H$ (or the self-adjoint operators $A=A^*$). In this case, the eigenvalues are real and the adjoint problem for the left eigenvectors reduces to $A \mathbf{u} = \lambda \mathbf{u}$ and $L^H=R^H$. It is therefore natural to take $D=I$ and $R$ as a unitary matrix $R^H R=I$. The Hermitian case is very comfortable since there is a simple one-one correspondence between the right and left eigenvectors by Hermitian conjugation. Indeed, these concepts are irrelevant in the Hermitian case and there is no ambiguity speaking about eigenvectors.

In the non-Hermitian case, the distinction between the right and left eigenvectors is fundamental and there is no way to build a left eigenvector from its corresponding right eigenvectors. The only possibility is to build all the left eigenvectors from all the right eigenvectors by inverting the $R$ matrix. This procedure is in fact intractable in practice. Consider a huge finite element discretisation: the larger eigenvalues just correspond to numerical noise and the computation of the explicit inverse would just be a numerical abomination. Usually, only a quite small subset of the eigenvectors are necessary for practical computations and the left eigenvectors will be computed by solving the eigenproblem provided by the Hermitian conjugate matrix.

\subsection{The linear operator pencils and the modal expansion\label{pencils}}

We consider now a generalized eigenvalue problem associated to the \gui{linear} matrix pencil $T(\lambda) = A - \lambda B$ (with $B$ a nonsingular matrix). A right eigenvector and its eigenvalue are defined by $T(\lambda)\mathbf{v} =A \mathbf{v} - \lambda B \mathbf{v} =0$ and a left eigenvector and its eigenvalue are defined by $\mathbf{u}^{H} T(\lambda) =\mathbf{u}^{H} A - \lambda \mathbf{u}^{H} B=0$, that is, using again $L^H$, $R$, and $\Lambda$ matrices, $AR= BR\Lambda$ and $L^HA=\Lambda L^H B$. Using an arbitrary nonsingular diagonal matrix $D$, we can write $D^{-1} L^H A B^{-1} L^{-H} D = R^{-1} B^{-1} A R = \Lambda$ and therefore: $$L^H B R= D,$$ that is a \textbf{$B$-weighted orthogonality.}

To build a modal expansion for the family of operators $T(z)= A -z B$, we consider the unknown vector $\mathbf{x}$, the given source vector $\mathbf{s}$, and a given complex number $z$ (the excitation frequency \gui{which is usually real}) such that: $$T(z) \mathbf{x} = (A - zB) \mathbf{x} = \mathbf{s}.$$
We consider a modal development $\mathbf{x} = \sum_i x_i \mathbf{v}_i $ with unknown complex coefficients $x_i$ and right eigenvectors $\mathbf{v}_i$ associated to eigenvalues $\lambda_i$. We multiply the equation on the left by a left eigenvector $\mathbf{u}_j^H$ to get: $\mathbf{u}_j^H T(z) \mathbf{x} = \mathbf{u}_j^H(A - zB) (\sum_i x_i \mathbf{v}_i)  = \mathbf{u}_j^H \mathbf{s}=$ $\sum_i x_i (\mathbf{u}_j^H A \mathbf{v}_i - z \mathbf{u}_j ^H B \mathbf{v}_i)$. We now use the fact that $A \mathbf{v}_i = \lambda_i B \mathbf{v}_i $ and the $B$-weighted biorthogonality $\mathbf{u}_j ^H B \mathbf{v}_i = \delta_{i,j} \mathbf{u}_j ^H B \mathbf{v}_j $ with $\delta_{i,j}$, the Kronecker symbol and we thus have: $$x_j(\lambda_j - z) \mathbf{u}_j ^H B \mathbf{v}_j = \mathbf{u}_j^H \mathbf{s}. $$
For any $z$ in the resolvent set ({\emph{i.e.}} not in the spectrum, $z \neq \lambda_i, \forall i $),
we can write $$\mathbf{x} = \sum_j \frac{1}{(\lambda_j - z)} \frac{\mathbf{u}_j^H \mathbf{s}}{\mathbf{u}_j ^H B \mathbf{v}_j }\mathbf{v}_j .$$
From this expression, we directly deduce the modal expansion of $T(z)^{-1}$, the \emph{resolvent} of $T(z)$ such that $\mathbf{x} = T(z)^{-1} \mathbf{s} $:
$$T(z)^{-1}=  \sum_j \frac{1}{(\lambda_j - z)} \frac{\mathbf{v}_j \mathbf{u}_j^H}{\mathbf{u}_j ^H B \mathbf{v}_j }. $$
The structure of this operator is probably more obvious using the bra-ket notation:
$$T(z)^{-1}=  \sum_j \frac{1}{(\lambda_j - z)} \frac{\ket{\mathbf{v}_j} \bra{\mathbf{u}_j}}{\braket{\mathbf{u}_j} {B \mathbf{v}_j} }. $$
See~\cite{laux2012solving} for applications of this formula in physics.

To have a flavor of modal expansion in the nonlinear eigenvalue problem case associated to a more general holomorphic $T(z)$ matrix, we can linearize this $T(z)$. Given a fixed $z_0$, we have $$T(z)= T(z_0) + (z-z_0) T'(z_0) + \mathcal{O}((z-z_0)^2) \approx  (T(z_0) - z_0 T'(z_0) )+ z T'(z_0) $$ and we have a local approximation of the resolvent using the expansion for a linear pencil with $A = T(z_0) - z_0 T'(z_0) $ and $B=-T'(z_0)$.
This approximation is in fact very close to the result provided by the Keldy\v{s} theorem introduced above.

\section{Time dispersion and rational interpolation}

One of the main practical questions is the representation of the frequency dependent coefficient involved in the $T(z)$ operator. As a paradigmatic example, we consider
 the time dispersive behaviour $\varepsilon(\omega)$ of the dispersive dielectric permittivity that is a function of the (angular) frequency $\omega$. A very simple model is that the electric field $\mathbf{E}$ and the electric polarisation $\mathbf{P}$ are locally related by an ordinary differential equation $\sum_k b_k \partial_t^k  \mathbf{E}=\sum_j a_j \partial_t^j  \mathbf{P}$. In the Fourier domain $\partial_t \rightarrow i \omega $, and the permittivity is therefore represented by a rational function of frequency:
$$ \mathbf{D} = \varepsilon \mathbf{E} = \varepsilon_0 \mathbf{E} + \mathbf{P} = \varepsilon_0 (1 + \chi(\omega) ) \mathbf{E} = ( \varepsilon_0 + \frac{\sum_j a_j (i\omega)^j }{\sum_k b_k(i\omega)^k}) \mathbf{E}.$$

 This representation is compatible with the most common models for dielectrics and metals in optics such as the Drude and Lorentz models and is a natural generalization of them. 
 
 It is also naturally causal with these rational functions made of real coefficients multiplying powers of $i \omega$.

{\crevii In system theory, rational approximation of frequency domain responses is a well established technique~\cite{GustavsenVectorFit} and a similar robust algorithm has been recently developed for electric permittivity that provides very accurate approximations on wide ranges of real frequencies in the optical domain from experimental measurements~\cite{Mauricio17permittivity,MauricioAlgo} and respecting the Kramers-Kronig constraints.}

 The obtained rational functions provide a natural analytical continuation of the media characteristics to the complex plane that is necessary to deal with complex eigenvalues characterizing the resonances of the photonic devices. Moreover, these analytical continuations unveil the electron resonances in the bulk material. It is worth noting that these real functions that are the trace of a complex meromorphic function on the real axis are very badly interpolated by polynomials  when the rational interpolation naturally locates the poles by analytic continuation.
 
\begin{figure}[ht!]
\centering
\includegraphics[width=0.4\textwidth]{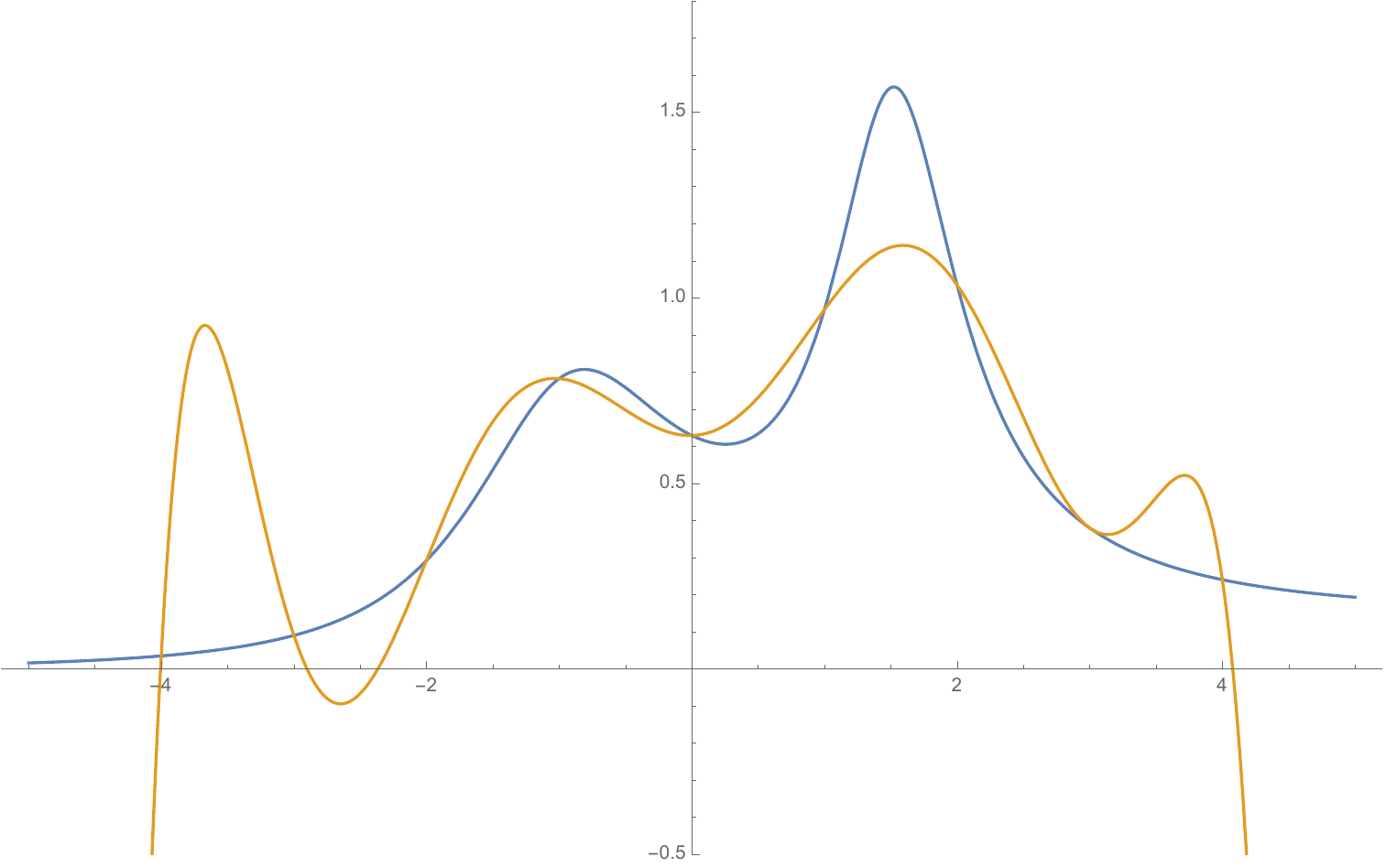}
\includegraphics[width=0.4\textwidth]{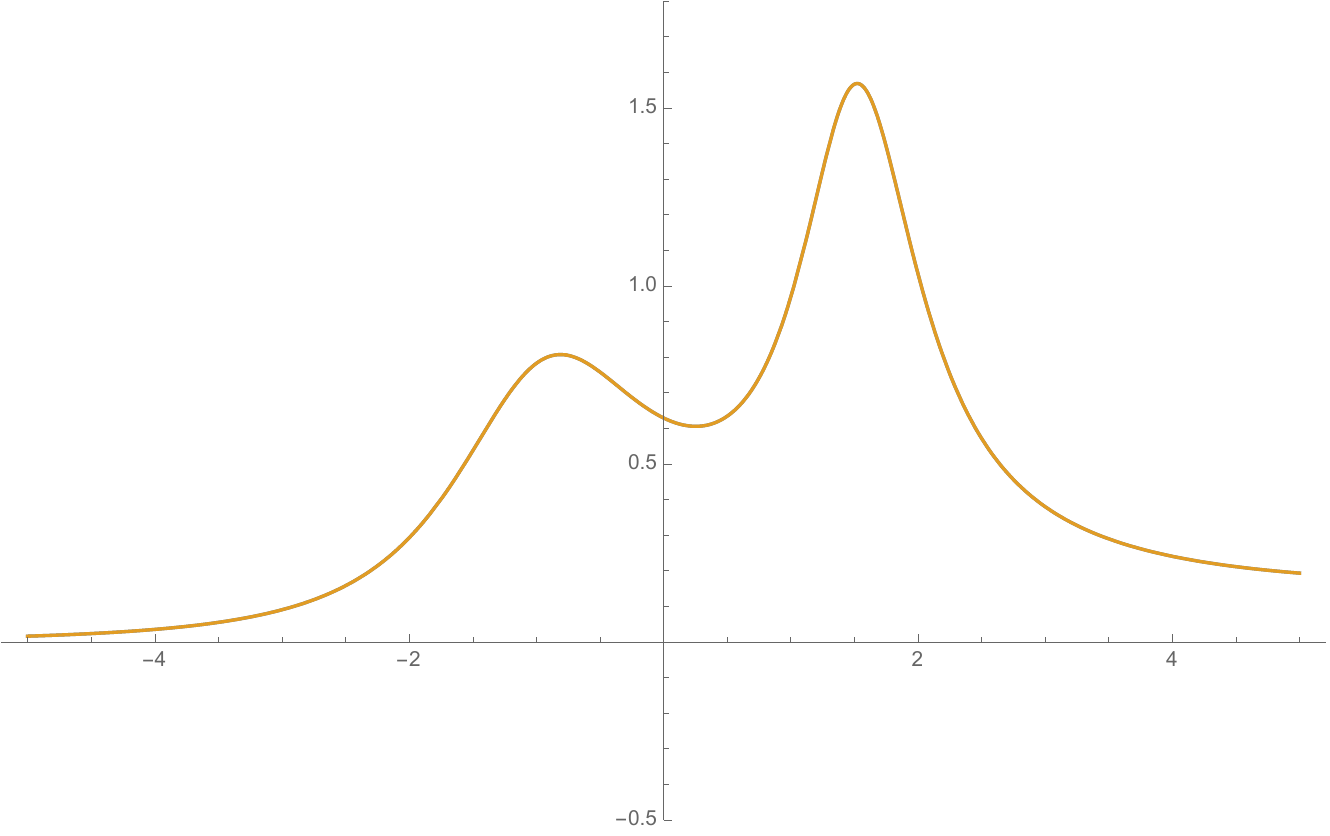}
\caption{Function $f(x)$ (blue curve) compared with its interpolation (orange curve) by the polynomial interpolation $f_{pol9}(x)$ (left) and by the rational interpolation $f_{rat7}(x)$ (right) for real values of $x$. }\label{cauplot}
\end{figure}

\begin{figure}[ht!]
\centering
\includegraphics[width=0.4\textwidth]{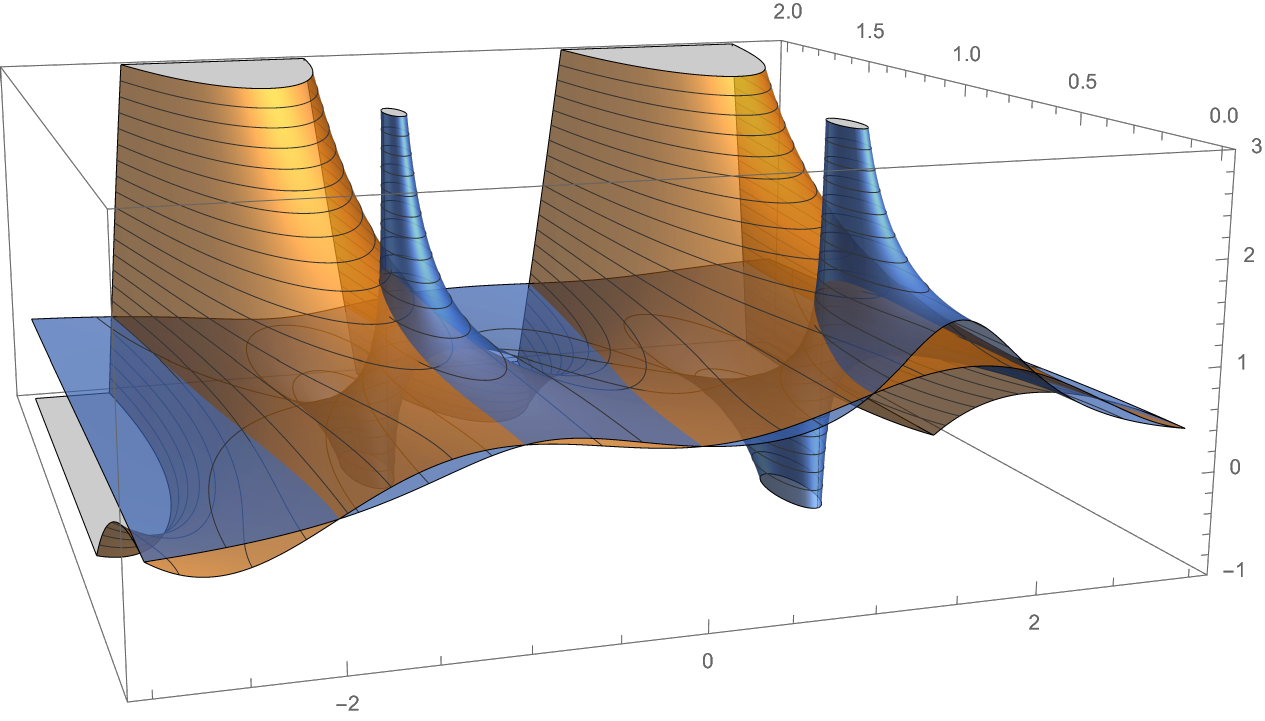}
\includegraphics[width=0.4\textwidth]{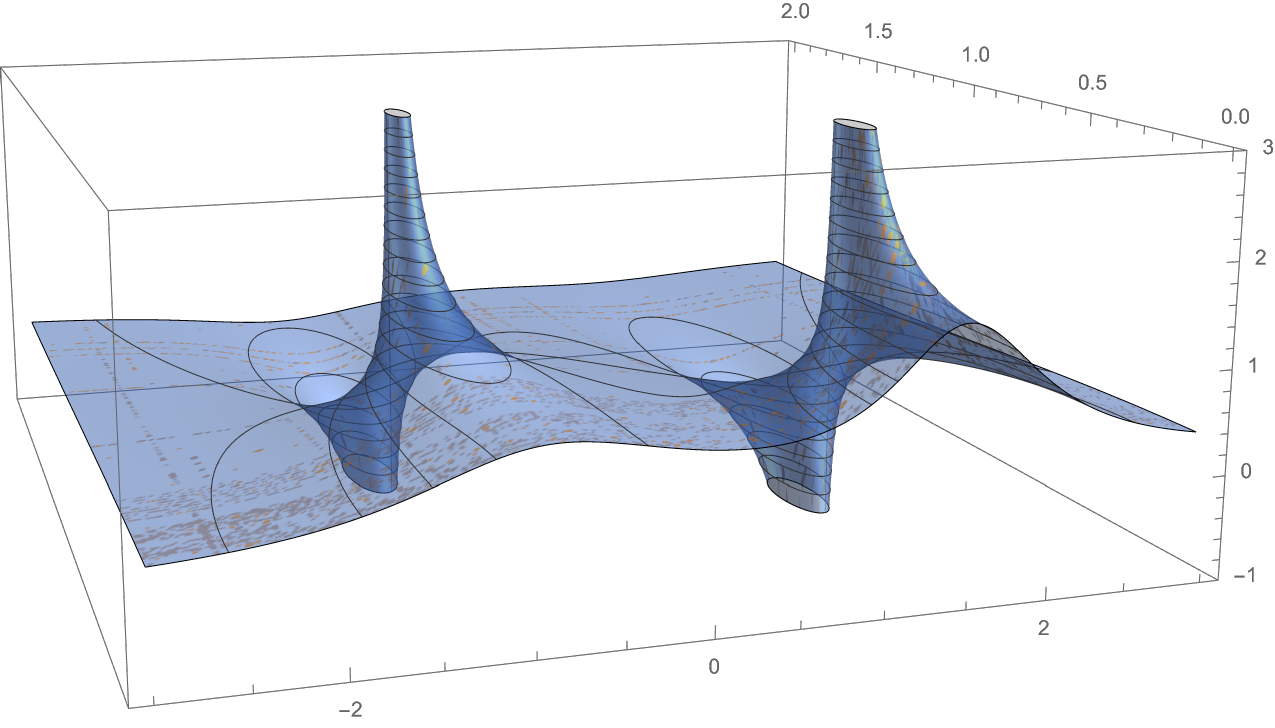}
\caption{Analytical continuation of function $f(x)$ (blue surface) in the positive imaginary part half complex plane containing two poles compared with its interpolation (orange surface) by the polynomial interpolation $f_{pol9}(x)$ (left) and by the rational interpolation $f_{rat7}(x)$ (right) for real values of $x$. }\label{cauplot3D}
\end{figure}

As an example, we consider the function $$f(x) = \dfrac{e^{0.3 x}}{(x+1)^2+1}+\dfrac{e^{0.2 x}}{3 (x-1.5)^2+1}$$ that has 4 complex poles but is not a rational function. The polynomial interpolation on 9 equally spaced points:
\begin{align*}
f_{pol9}(x) = & -0.000503751 x^8+0.000621437 x^7+0.0156374 x^6-0.0160013 x^5-0.146763 x^4 \\ 
& + 0.0978923 x^3+0.377692 x^2+0.0103092 x+0.629032
\end{align*}
is completely wrong while rational interpolation on 7 points only $$f_{rat7}(x) = \dfrac{-0.0844279 x^3-0.0745097 x^2+0.992407 x+0.629032}{0.0840486 x^3-0.470427
   x^2+0.897358 x+1}$$ gives a perfect result  (Fig. \ref{cauplot}). Moreover, in practice, the measurements performed to identify a transfer function or a material property are made with real frequencies, but in the context of our dispersive model, we need the analytical continuation of the measured quantity. 
Materials properties often turn out to have intrinsic resonances associated to poles in the corresponding transfer function. The polynomial is irremediably without any poles while the rational interpolation is matching the poles of the original function as shown on Fig. \ref{cauplot3D}.

 The degrees of the numerator and denominator are chosen to fit the data but there are also physical constraints. For instance, it can be shown~\cite{lucarini2005kramers} that at very high frequencies, the electric susceptibility  $\chi(\omega)$ decreases asymptotically as $\dfrac{1}{ \omega^2}$.
 
 Another advantage of the rational representation is that it is well suited to nonlinear eigenvalue algorithms~\cite{Hernandez:2005:SSF}.

The rational (causal) interpolation from measurements at real frequencies and its straightforward analytical continuation is a model that can certainly be applied to a lot of medium characteristics in various fields of physics.

\section{Polynomial matrix eigenvalue problems}
\label{sec:polynomial}

In this section we determine the modal expansion of a polynomial dispersive operator $T(z)$, that is a matrix with coefficients that are polynomials of $z$, from its nonlinear eigentriplets.

In the sequel, in order to simplify notations, the $\mathbf{v}$'s will denote column vectors and the $\mathbf{w}$'s row vectors (here, we give up the explicit $^H$ superscript since there is no ambiguity and the corresponding column vector is never used).

The polynomial eigenvalue problem can be put in the form $\mathbf{A}(\lambda_i) \mathbf{v}_i = (\sum_{k=0}^m \lambda_i^k \mathbf{L}_k) \mathbf{v}_i=\mathbf{0}$ where $m$ is the maximum degree of the polynomials involved in the coefficients, $\lambda_i$ is an eigenvalue and $\mathbf{v}_i$ the corresponding $n$-dimensional (right) eigenvector ($n$ is the number of complex parameters in the discrete model). The $\mathbf{L}_k$  are $n \times n$ matrices with constant complex coefficients. The eigenvalue problem can be linearized using the $n m \times n m$  \textbf{companion matrix} $\mathcal{A}$ (also used to turn high order differential equations into a system of first order equations) acting on an $n m$ (column) vector:
\textbf{}
 $$\mathcal{V}_\lambda(\mathbf{v})=\left(\begin{array}{c}\mathbf{v} \\ \lambda \mathbf{v} \\ \vdots \\ \lambda^{m-1} \mathbf{v} \end{array}\right).$$
 
 The corresponding linearized eigenvalue problem $\mathcal{A} \mathcal{V}_\lambda(\mathbf{v })=  \lambda \mathcal{V}_\lambda(\mathbf{v })$ provides a set of $n m$ eigenpairs $( \mathcal{V}_{\lambda_i}(\mathbf{v_i }), \lambda_i )$. 

In practice, we are interested in simply reconstructing $\mathbf{v}$, that corresponds to the physical field, rather than $\mathcal{V}_{\lambda}(\mathbf{v})$ that contains a redundant information since it is entirely determined by a $(\lambda, \mathbf{v})$ pair. The vector space of the vectors $\mathbf{v}$ is of dimension $n$ and the $\mathcal{V}_{\lambda}(\mathbf{v})$ are elements of an $n m$ dimension vector space, hence they are too many for the dimension $n$ space and the reconstruction of a $\mathbf{v}$ vector is necessarily not unique as it appears explicitly below.

 In the non-Hermitian case, we also need the (left) eigenvectors such that $\mathbf{w}_i\mathbf{A}(\lambda_i)=\mathbf{0}$. Considering the (row) vectors $\mathcal{W}_{\lambda_i}(\mathbf{w}_i)=(\mathbf{w}_i \; \lambda_i \mathbf{w}_i \; \cdots  \; \lambda_i^{m-1} \mathbf{w}_i )$ and using the companion matrix $\mathcal{A}$, we have the equivalent linearized left eigenvalue problem  $\mathcal{W}_{\lambda_i}(\mathbf{w_i })\mathcal{A}=  \lambda_i \mathcal{W}_{\lambda_i}(\mathbf{w_i })$. 
  
Finally, eigentriplets $(\lambda_i,\mathcal{W}_{\lambda_i}, \mathcal{V}_{\lambda_i} )$ are obtained
that can be used for a modal expansion (we do not consider here the case of defective matrices) of $\mathcal{A}^{-1}$.   
Nevertheless, the interesting result would be to express $\mathbf{A}(\lambda)^{-1}$ in terms of the (``small'') $n$-dimensional vectors $\mathbf{v}_i$ and  $\mathbf{w}_i$. This computation has been performed by Lancaster \cite{lancaster1960inversion, lancaster1966lambda} sixty years ago for the study of mechanical vibrations and gives:

$$\mathbf{A}(\lambda)^{-1} = \sum_{i=1}^{n m}  \frac{\lambda_i^r}{\lambda^r} \frac{1}{\lambda-\lambda_i} \frac{ \mathbf{v}_i  \mathbf{w}_i}{\mathbf{w}_i \mathbf{A}'(\lambda_i) \mathbf{v}_i} \quad  \mathrm{for} \: r= 0, \cdots, m-1$$
where $\mathbf{v}_i   \mathbf{w}_i$
is 
the matrix product between the column right eigenvector $\mathbf{v}_i $ and the row left eigenvector $\mathbf{w}_i $ resulting in a square matrix, $\mathbf{A}'(\lambda_i) = d\mathbf{A}(z)/dz\mid_{z=\lambda_i}$ is the complex derivative of $\mathbf{A}(\lambda)$ (obtained by taking the complex derivatives of the coefficients) evaluated at $ \lambda=\lambda_i$.
We can see that we have $m$ different expansions corresponding to the possible choices for $r$ and that for $r=0$, we recover the particular case corresponding to the application of the Keldy\v{s} theorem~\cite{zolla2018photonics}. It is quite easy to see that these expansions can be linearly combined to replace the factors $ \dfrac{\lambda_i^r}{\lambda^r}$ by $ \dfrac{p(\lambda_i)}{p(\lambda)}$ where $p(x)$ is an arbitrary polynomial of degree up to $m-1$.

\section{Rational matrix eigenvalue problems}
\label{sec:rational}

In order to introduce the case with rational coefficients,
we generalize the polynomial case by considering a linear combination of the $\mathbf{L}_k$ constant matrices with scalar coefficients that are now rational functions of the frequency instead of simple powers: $\mathbf{A}(\lambda_i) \mathbf{v}_i = (\sum_{k=1}^m R_k(\lambda_i) \mathbf{L}_k) \mathbf{v}_i=\mathbf{0}$. We define $R_k(\lambda) = \dfrac{N_k(\lambda) }{D_k(\lambda) } $ with $N_k$ the numerator polynomial of degree $n_k$ and $D_k$ the denominator polynomial of degree $d_k$. We define $\mathbf{P}_k(\lambda)=N_k(\lambda) \mathbf{L}_k $, $D(\lambda)=\prod_{i = 1}^m D_i(\lambda)$ (polynomial of degree $d = \sum d_i$) and $\hat{D}_k(\lambda)=\prod_{\substack{i=1 \\ i \neq k}}^m D_i(\lambda)$ (polynomials of degrees $d - d_k$).

Multiplying by $D(\lambda)$, we turn the rational problem into a polynomial one: 
$D(\lambda) \mathbf{A}(\lambda) = \sum_{k=1}^m \hat{D}_k(\lambda) N_k(\lambda) \mathbf{L}_k  = \sum_{k=1}^m \hat{D}_k(\lambda) \mathbf{P}_k(\lambda)$ with a degree $m=\max(n_k-d_k +d)$.

We still consider the, now rational, nonlinear eigenvalue problem $\mathbf{A}(\lambda_i) \mathbf{v}_i =\mathbf{0}$ (and the corresponding left one). There are currently very efficient algorithms to solve them, as discussed in section~\ref{sec:numerical}.
The corresponding polynomial problem $D(\lambda_i) \mathbf{A}(\lambda_i) \mathbf{v}_i =\mathbf{0}$ shares the same eigenvalues and eigenvectors. Indeed we suppose here that the $\lambda_i$ are not roots of $D(\lambda)$. These roots correspond to bulk material resonances and play of course an important role in the physical systems under study. The roots of the $N_k(\lambda)$'s, corresponding $e.g.$ to null permittivity, may play an important role too but these questions do not interfere here. We now apply the polynomial case expansion:
$$(D(\lambda) \mathbf{A}(\lambda))^{-1} = \sum_{i=1}^{n m}  \dfrac{p(\lambda_i)}{p(\lambda)} \frac{1}{\lambda-\lambda_i} \frac{ \mathbf{v}_i  \mathbf{w}_i}{\mathbf{w}_i (D\mathbf{A})'(\lambda_i) \mathbf{v}_i}.$$
We have: $(D\mathbf{A})'(\lambda_i)\mathbf{v}_i=$
$D'(\lambda_i)\mathbf{A}(\lambda_i)\mathbf{v}_i+D(\lambda_i)\mathbf{A}'(\lambda_i)\mathbf{v}_i=$ 
$D(\lambda_i)\mathbf{A}'(\lambda_i) \mathbf{v}_i$. 
So we have:
$$\mathbf{A}(\lambda)^{-1} = \sum_{i=1}^{n m}  \dfrac{p(\lambda_i)D(\lambda)}{p(\lambda)D(\lambda_i)} \frac{1}{\lambda-\lambda_i} \frac{ \mathbf{v}_i  \mathbf{w}_i}{\mathbf{w}_i \mathbf{A}'(\lambda_i) \mathbf{v}_i}.$$
This formula is very general and can be used in practice \cite{truong2020continuous} but a simpler form will be worked out without any loss of generality for all practical purposes.
In order to obtain a simple form similar to the one of the polynomial problem, we take $p$ to be a multiple of $D(\lambda)$ $i.e.$ $p(\lambda) = q(\lambda) D(\lambda) $. The conditions of existence of the quotient polynomial $q(\lambda)$ is that the degree of $p$ is larger than or equal to the degree of $D(\lambda)$: $m-1=\max(n_k-d_k +d)-1 \leq d  $ that is $\max(n_k-d_k ) \leq 1$. A sufficient condition is that at least one of the $D_k$ has $n_k > d_k$ and the degree of $q(\lambda)$ $= n_k -d_k-1 \geq 0$. Finally, we have:

$$\mathbf{A}(\lambda)^{-1} = \sum_{i=1}^{n m}  \dfrac{q(\lambda_i)}{q(\lambda)} \frac{1}{\lambda-\lambda_i} \frac{ \mathbf{v}_i  \mathbf{w}_i}{\mathbf{w}_i \mathbf{A}'(\lambda_i) \mathbf{v}_i}.$$

In the case of Maxwell's equations with dispersive permittivity, the relevant term in the electric field wave equation is $\omega^2 \varepsilon(\omega)$. We have seen that the susceptibility vanishes at very high frequencies and the permittivity tends to a constant. It is therefore represented by a rational function with the numerator and the denominator of the same degree. Finally, considering the rational function for $\omega^2 \varepsilon(\omega)$, we have $n_k - d_k -1 = 1$. In this case, $q(\lambda) = a \lambda + b$. We have a one parameter family of expansions:
$\dfrac{q(\lambda_i)}{q(\lambda)} = \dfrac{ \lambda_i + c}{\lambda + c}$, $c \in \C$ with the notable cases $c=0 \Rightarrow \dfrac{q(\lambda_i)}{q(\lambda)} = \dfrac{\lambda_i}{\lambda}$, $c\rightarrow \infty \Rightarrow  \dfrac{q(\lambda_i)}{q(\lambda)}=1$, and $c=-\lambda_i \Rightarrow  \dfrac{q(\lambda_i)}{q(\lambda)}=0$ that remarkably cancels the contribution of a particular resonance in the expansion~\cite{truong2020continuous}.

\section{Numerical computation of eigenvalues and eigenvectors}
\label{sec:numerical}
The theory discussed in the previous two sections relies on the computation of the eigentriplets $(\lambda_i, \mathbf{w}_i, \mathbf{v}_i )$ of either a polynomial or rational eigenvalue problem. But care must be taken when doing the computation in practice, since numerical difficulties may arise. In this section, we point out several issues to consider, all of which have been addressed in the implementation of solvers included in the SLEPc library~\cite{Hernandez:2005:SSF} that we use for the results in next section.

The standard approach for solving polynomial eigenvalue problems is the linearization via the companion matrix that was discussed in section~\ref{sec:polynomial}. As mentioned, the linearization implies an increase of the dimension by a factor $m$ (the degree of the polynomial), which translates into much higher memory requirements and computational cost. This issue can be avoided by using a compact representation of the large subspace basis, expressed in terms of tensor products of a basis of the small subspace~\cite{Campos:2016:PKS}. A robust numerical implementation must also apply some kind of scaling to equilibrate the norms of the coefficients $\mathbf{L}_k$ in case they differ wildly~\cite{Campos:2016:PKS}.

Another pitfall of polynomial eigensolvers based on linearization is that, in the case of high degree $m$, the representation of the polynomial in terms of monomials, i.e., the successive powers of $\lambda$, is not appropriate since the vectors $\mathcal{V}_{\lambda}(\mathbf{v})$ are then defined by scaling factors of very different magnitude, which leads to numerical instability. It is possible to replace the monomial basis by a more appropriate polynomial basis such as Chebyshev and others~\cite{Campos:2016:PKS}, but each of them is appropriate only for the case that the wanted eigenvalues are located in certain parts of the complex plane.

An alternative to polynomial eigensolvers based on linearization are eigensolvers that operate directly on the original space, such as Jacobi-Davidson~\cite{Campos:2020:PJS} or contour integral methods.

Regarding the rational eigenproblem discussed in section~\ref{sec:rational}, even though it is feasible to transform the problem to a polynomial eigenproblem, the recommendation is not to proceed this way, to avoid handling a high-degree polynomial. Instead, the rational matrix $\mathbf{A}(\lambda)$ can be linearized directly, in a similar way to polynomial matrices, as is done by the NLEIGS method~\cite{guttel2017nonlinear,Campos:2016:PKS}. More precisely, NLEIGS is an algorithm intended for general nonlinear matrix-valued functions (not only rational), that are first approximated by a rational matrix via interpolation and then the rational matrix is linearized. In the case that the input matrix-valued function is already rational, the interpolation produces an equivalent rational representation, with the benefit that scaling factors are introduced to avoid significant differences in matrix norms. The dimension of the linearization is again $m$ times larger, where $m$ is now the value defined in section~\ref{sec:rational}. The implementation of NLEIGS in SLEPc~\cite{Campos:2016:PKS} was used for the results shown in the next section.

The last comment is about computation of left eigenvectors $\mathbf{w}_i$. In SLEPc, not all eigensolvers are prepared to compute both right and left eigenvectors, in which case one should have to solve both the direct problem and its adjoint. Luckily, SLEPc's NLEIGS solver is implemented in a way that both sets of eigenvectors are returned, even though this implies giving up the compact representation and using a full basis for the large-dimensional space~\cite{Campos:2016:PKS}.

SLEPc's solver objects hold the computed eigenvectors internally, and when the user requests to multiply the resolvent by a vector it does so by implicitly applying the resolvent formulas discussed previously, without forming the resolvent matrix explicitly. The accuracy of the result will depend on how many eigentriplets have been computed, since the resolvent expansion will include only those.

\section{The physically agnostic expansion}
\subsection{A 2D example}
\begin{figure}[ht]
   \centering
   \includegraphics[width=\textwidth]{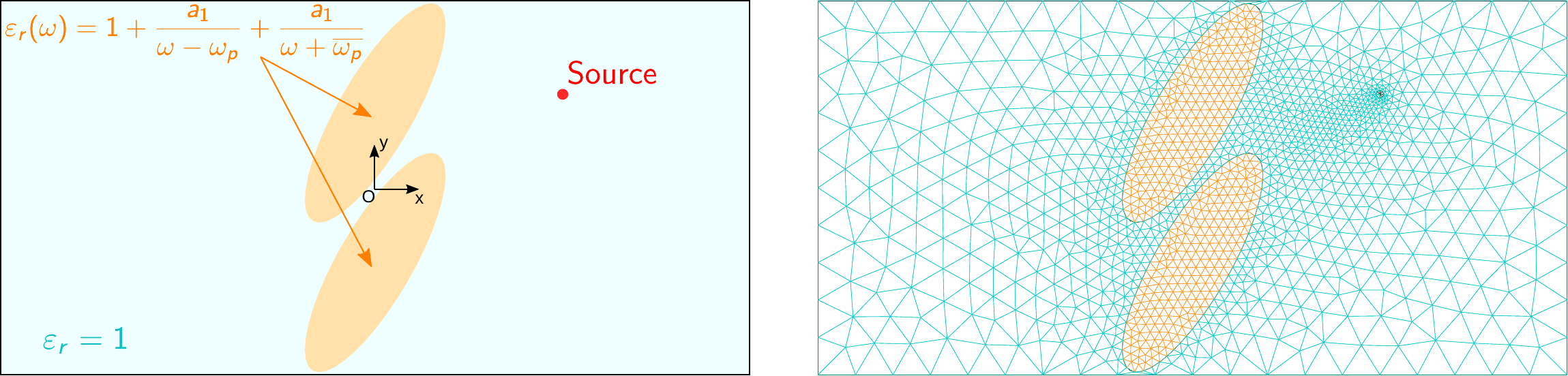}
   \caption{A rectangular box of size $2L\times L$ enclosing two dispersive ellipses in a vacuum. A source is located in $(L/2,L/4)$.}
   \label{fig:scheme}
\end{figure}

Now let us consider a practical example and its numerical solution using (i) the direct implementation, (ii) a physical expansion and (iii) the ``physically agnostic'' expansion considered above. For instance, a rectangular perfectly conducting cavity is assumed to contain two dispersive ellipses and a simple oriented delta as a source (see left panel in Fig.~\ref{fig:scheme}). The TM case ($E_z=0$) is solved in this example, and we choose the vector unknown $\mathbf{E}=(E_x,E_y,0)$ as this case exhibiting surface plasmon and components discontinuities is more representative of the 3D case. The illustrative goal is to obtain the absorption spectrum, \emph{i.e.} losses inside the dispersive media as a function of the source frequency denoted $\omegas$. The traditional approach when using frequency domain FEM is to loop over the collection of direct problems resulting from the frequency sweeping in the vector propagation equation: 
$-\Curl \, \Curl \, \mathbf{E} + (\omegas/c)^2 \varepsilon_r(\mathbf{r},\omegas) \, \mathbf{E} =\delta(x_S,y_S)\mathbf{x}.$

The relative permittivity $\varepsilon_r(\mathbf{r},\omega)$ of the problem is a function defined by parts, and it is ruled by a Lorentz model $\varepsilon_r^d$ inside the dispersive ellipses

$$  \varepsilon_r^d(\omega)=1+\frac{a_1}{\omega-\omega_p} + \frac{a_1}{\omega+\overline{\omega}_p}
$$ 
and equal to 1 outside. The value of the real part of the pole $\omega_p$ of the permittivity is chosen to be of the order of magnitude of the fundamental eigenfrequency of the bare cavity:  $\omega_p/\eta=0.3-0.025i$ and $a_1/\eta=-0.3$ with $\eta=4\pi c/L$.

A classical discretisation (see the mesh in the right panel of Fig.~\ref{fig:scheme}) using edge elements~\cite{dillon_webb} leads to the following matrix problem:
$$
c^2\mathbf{L}_1\mathbf{e}+\lambdas^2\mathbf{L}_2\mathbf{e}+\lambdas^2 \varepsilon_r^d(-i\lambdas)\mathbf{L}_3\mathbf{e}=\mathbf{p},
$$
where $\mathbf{e}$ is the unknown, $\mathbf{L}_1$ is the stiffness matrix resulting from the $\Curl \, \Curl$ term, $\mathbf{L}_2$ and $\mathbf{L}_3$ are two mass matrices assembled in each  material inside the box (freespace and the dispersive media respectively) and $\mathbf{p}$ is the load vector corresponding to the source. Note that the frequency dependence is factorized and solely appears in front of the discrete operators.

\begin{figure}[ht]
   \centering
   \includegraphics[width=\textwidth]{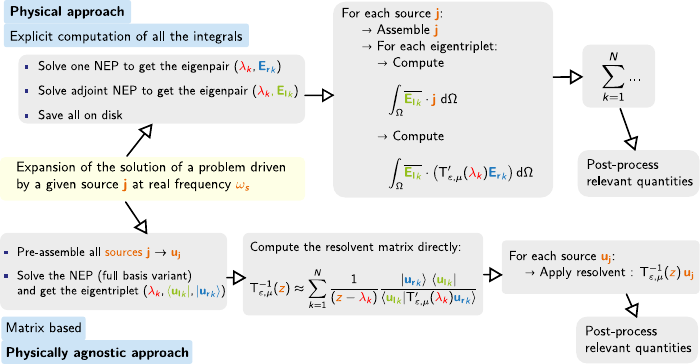}
   \caption{Summary of the two algorithms}
   \label{fig:algo}
\end{figure}

The source free companion problem ($\mathbf{p}=0$) forms a rational eigenvalue problem as introduced at the beginning of the previous section: $\mathbf{A}(\lambda_i) \mathbf{v}_i = (\sum_{k=1}^m R_k(\lambda_i) \mathbf{L}_k) \mathbf{v}_i=\mathbf{0}$, where $R_1=c^2$, $R_2=\lambda_s^2$ and $R_3=\lambda_s^2\varepsilon_r^d(-i\lambda_s)$.

Using the last expression of the resolvent found in the previous section, we now know that the solution of our problem subject to a particular load vector $\mathbf{p}$ at the real frequency $\omegas$ can be expanded as:
$$\mathbf{e}=\mathbf{A}(\lambdas)^{-1}\mathbf{p}.$$

The corresponding ``physical'' counterpart (\emph{i.e.} at the continuous level and integral based) of the expansion would be~\cite{truong2020continuous}:
$$
   \mathbf{E}=  \sum_{k=1}^{n} \frac{1}{(\lambdas-\lambda_k)} \frac{ \int_{\Omega} \overline{\mathbf{E_l}}_k \cdot (\delta(x_S,y_S)\mathbf{x}) \,d\Omega}{\int_\Omega \overline{\mathbf{E_l}}_k \cdot \left(  \TE'(\lambda_k) \mathbf{E_r}_k \right ) \,d\Omega} \mathbf{E_r}_k.
$$

We have implemented two strategies based on the two above formulas:
\begin{itemize}
   \item The ``physical'' one consists in several FEM runs (see the top path of Fig.~\ref{fig:algo}):
   \begin{itemize}
      \item solving the left and right rational eigenvalue problems \cite{Campos:2019:NEP} one by one and saving the eigenresults,
      \item computing all the overlap integral between the source and the left eigenvector in the previous expression,
      \item computing the all the integrals at the denominator of the previous expression,
      \item postprocess the sum. 
   \end{itemize}
   \item The ``physically agnostic'' remains at the matrix level and consists in (see the bottom path of Fig.~\ref{fig:algo}):
   \begin{itemize}
      \item computing eigentriplets of the rational eigenvalue problem at once using a two-sided solver from SLEPc,
      \item deducing the resolvent matrix directly also via SLEPc,
      \item applying the resolvent to all the pre-assembled source.
   \end{itemize}
\end{itemize}
In practice, the  open source Finite Element package ONELAB/Gmsh/GetDP (\texttt{http://onelab.info})~\cite{Dular-GetDP,geuzaine2020three} was used for the numerical experiments. The support of polynomial and rational eigenvalue problems through the SLEPc library \cite{Hernandez:2005:SSF,Campos:2019:NEP} have recently been introduced  \cite{demesy2020non} in GetDP and even more recently we have introduced the support of the numerical resolvent-based expansions.

\begin{figure}[ht]
   \centering
   \includegraphics[width=\textwidth]{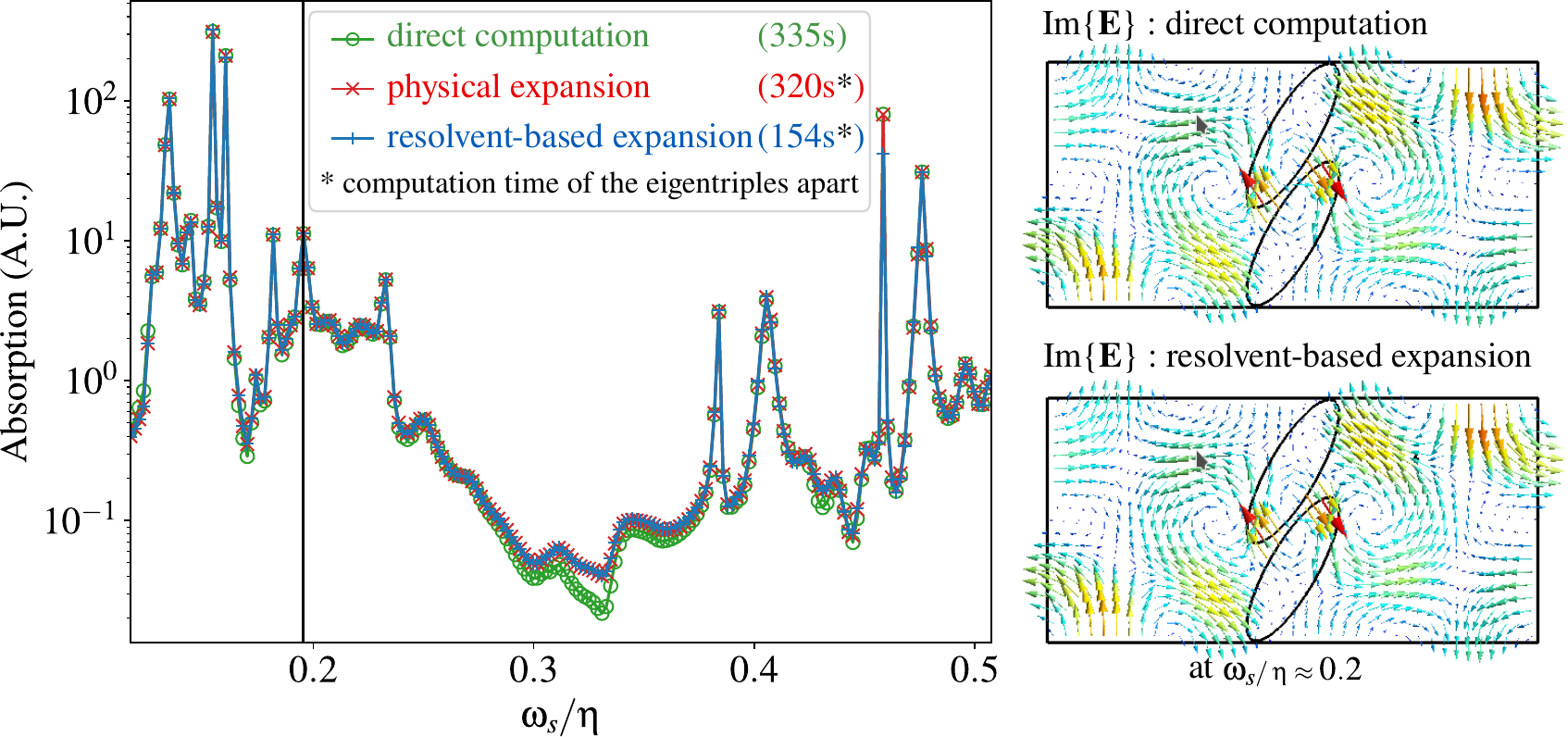}
   \caption{(Left) Absorption spectrum of the dispersive ellipses in log scale for the three approaches: direct computation in green color, ``physical'' integral-based expansion in red color and the ``physically agnostic'' resolvent-based expansion in blue color. (Right) Imaginary part of the electric field at $\omegas/\eta\approx0.2$ for the direct case (top) and the expanded case (bottom).}
   \label{fig:spectrum}
\end{figure}

Coming back to our example, the two expansion schemes are compared to the direct problem. The number of eigenvalues computed for the expansions is 150 while the number of requested real driving frequencies is 200. {\crevii The NLEIGS solver requires for the computation of eigenvalues to specify a rectangular region of interest in the complex plane and a target within this region. In this 2D closed case, the imaginary part of the spectrum is known to be bounded \cite{brule2016calculation} and lies in [Im$\{\omega_p/\eta\},0$], which gives the imaginary part interval of interest. Note that there are two accumulation points of eigenvalues with imaginary parts Im$\{\omega_p/\eta\}$ with corresponding eigenvectors having spatial frequencies tending towards 0, which cannot be captured by any mesh, so in practice the imaginary part interval of interest is chosen to be [0.8\,Im$\{\omega_p/\eta\},0$]. As for the real part interval, it should be chosen slightly larger than the real frequency range of interest}. The real part of the pole $\omega_p$ of the dispersive permittivity $\varepsilon_r^d$ is centered in the range of these driving frequencies. In other words, the expansion is performed in an ultra-dispersive frequency regime. We are facing here a structure exhibiting both strong interacting material and geometric resonances. The left part of Fig.~\ref{fig:spectrum} shows in log scale the Joule absorption inside the dispersive rods for the direct computation (green curve), the "physical" integral-based expansion (red) and the "physically agnostic" numerical resolvent-based expansion (blue). The agreement between the three results is excellent. The only visible discrepancy occurs  around $\omegas/\eta\approx0.3$ which corresponds to the real part of the pole $\omega_p$ of the dispersive permittivity: Indeed at this complex frequency the permittivity becomes infinite, and we expect there an infinite accumulation of eigenvalues which obviously one cannot reach numerically~\cite{zolla2018photonics,demesy2020non}. The hypotheses of the Keldy\v{s} theorem stated in Sec.~\ref{part:Keldysh} are only partly fulfilled since one cannot compute all the eigenvalues inside a given contour in the complex plane. Yet, this discrepancy can be witnessed in log scale only and the reconstruction of the field by the expansions is excellent as shown for a particular frequency in the right part of Fig.~\ref{fig:spectrum}.  

The computation times for performing the 200 expansions are given in the label of Fig.~\ref{fig:spectrum}. {\crevi This 2D case leads to 12000 degrees of freedom, leading to 7~Gb RAM usage for the resolvent based expansion. All the numerical examples in this paper were performed on a workstation equipped with Intel Xeon 8180M @ 2.5~GHz processors.} In spite of the fact that this case is very favorable to the ``physical'' integral-based expansion since the numerical support of overlap integrals between the source and the eigenvectors reduces to one single edge of the mesh, the ``physically'' agnostic resolvent based expansion time is  twice faster. Both expansion-based approaches are faster than the direct simulations, {\crevii but we excluded the computation time of the eigenvalues themselves which took 600~s in this case. Moreover, as will be shown in the next section, this last point does not remain true for larger 3D models.}

\subsection{\crevi{A 3D example}}
\begin{figure}[ht]
   \centering
   \includegraphics[width=0.9\textwidth]{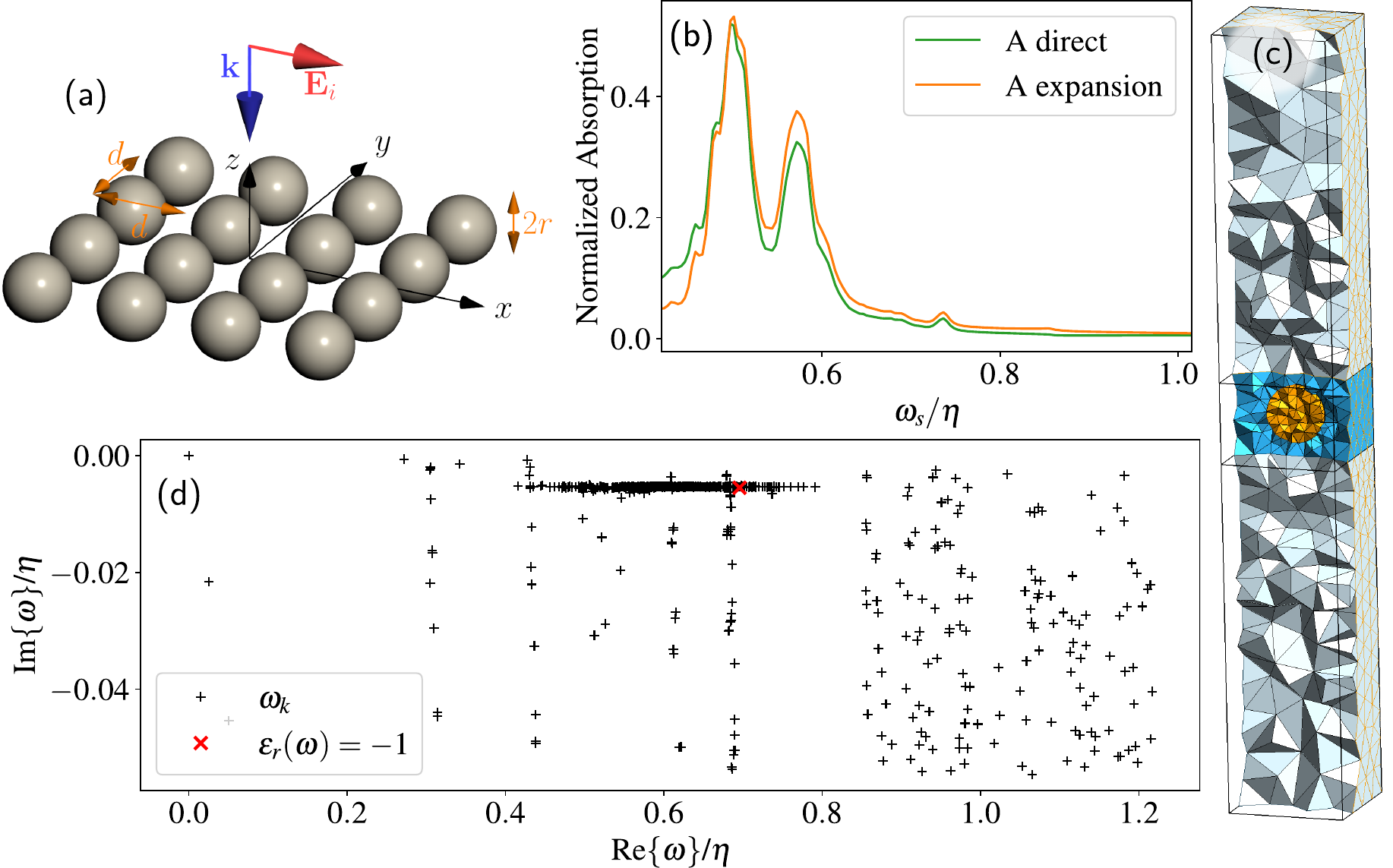}
   \caption{\crevi (a) Sketch of the 3D structure. (b) Absorption spectrum. Green: direct computation. Orange: expansion. (c) Cut of the mesh of a unit cell used in the computation. The tetrahedral elements constituting the dispersive sphere are depicted in orange color, those for the surrounding air in blue color and those for the PMLs in grey color. (d) Spectrum of complex eigenfrequencies of the structure (black pluses). The red cross shows the location of the plasmonic accumulation point.}
   \label{fig:grating3D}
\end{figure}

{\crevi After this convincing yet rather academic validation, we consider a realistic 3D example: a square bi-periodic 3D structure made of frequency dispersive objects illuminated by a plane wave as depicted in Fig.~\ref{fig:grating3D}(a).}

{\crevi The objects are spheres arranged in a square lattice with lattice constant $d$. We are considering the Joule absorption by such an open structure as it is illuminated by a plane wave of wavevector $\mathbf{k}=[0,0,-\omegas/c]^T$ polarized along $(Ox)$. The incident electric field writes $\mathbf{E}_{i} = [\exp(-iz\omegas/c ),0,0]^T$. As in the previous example, the relative permittivity $\varepsilon_r(\mathbf{r},\omega)$ of the problem is a function defined by parts, and it is ruled by a Drude model $\varepsilon_r^d(\omega)$ inside the dispersive spheres: $$  \varepsilon_r^d(\omega)=1-\frac{\omega_d^2}{\omega\,(\omega-i\gamma_d)} $$ and equal to 1 outside. Bearing in mind the normalization factor $\eta=2\pi c/d$, the relevant numerical values used in this example are the radius of the spheres $r=d/2$, the plasma frequency $\omega_d=1.62/\eta$ and the damping $\gamma_d=0.02/\eta$. Choosing $d$ around 0.5~$\mu$m leads to typical problems \cite{collin2010nearly,Lalanne:19} encountered in optics -- refered to as plasmonic problems -- where the real part of the permittivity of the objects becomes negative.

The traditional route to compute the absorption spectrum of such a structure consists in a frequency sweeping in the vector propagation equation to determine for instance the scattered field defined as $\mathbf{E}_s= \mathbf{E}-\mathbf{E}_i$ and solution of  : 
$-\Curl \, \Curl \, \mathbf{E}_s + (\omegas/c)^2\,\varepsilon_r(\mathbf{r},\omegas) \, \mathbf{E}_s = (\omegas/c)^2\,(1-\varepsilon_r(\mathbf{r},\omegas))\,\mathbf{E}_i.$
Note that the RHS term is far less trivial than in the previous example since it now depends on both space and $\omegas$. It is bounded inside the spherical objects. Bloch quasi-periodic conditions are used to bound the computational domain along the periodic directions $(Ox)$ and $(Oy)$ while Perfectly Matched Layers are used along the infinite direction $(Oz)$. The absorption spectrum  $A(\omegas)$,  normalized to the incident power, is postprocessed by computing the following quantity $A(\omegas)=\omegas\varepsilon_0\mathrm{Im}\{\varepsilon_r^d(\omegas)\}/(2P_i)\int_{\Omega_s}|E|^2\,\mathrm{d}\Omega$, where $P_i$ is the incident power through a unit cell. The absorption obtained using a direct conventional FEM method is shown in green color in Fig.~\ref{fig:grating3D}(b) while the absorption obtained after expansion (using the 750 modes shown in Fig.~\ref{fig:grating3D}(c)) is shown in orange color.

The agreement between the two methods in this frequency range is satisfactory, but it was nearly perfect in the previous academic 2D example. The main reason accounting for this performance drop is the openness of the structure, \emph{i.e.} the periodic structure couples to the freespace continuum of modes. This is modelled by PMLs. The theoretical spectrum of a PML medium is the real axis rotated by an angle corresponding to the prescribed damping. For a numerical finite thickness PML, this line becomes discrete \cite{vial2014quasimodal} so that the imaginary part of the eigenvalues of such structures is no longer bounded as in the previous 2D academic case, which makes the spectrum of the structure unbounded as well. As a consequence, there exists an infinity of modes with associated eigenvalue away from the region of frequency of interest that still contribute to the response of the structure.

However, the so-called plasmonic accumulation point of eigenfrequencies ($\varepsilon_r^d(\omega)=-1$, see the red marker in Fig.~\ref{fig:grating3D}(c)) allows to explain \cite{Lalanne:19} the shape of the absorption spectrum more that qualitatively. For this model, full second order edge element where used (40000 degrees of freedom), the computation time for computing the direct problems (120 points) is 686~s. The ``physical'' integral-based expansion took 5863~s while the ``physically'' agnostic resolvent-based expansion took 2470~s (with 37Gb of RAM memory usage). Compared to the 2D case, the matrices considered here are much larger and less sparse due to the higher connectivity of 3D elements. The computation times observed are different: the direct problem is now faster to run {\crevii (using the direct solver MUMPS \cite{MUMPS:1})}, but the ``physically'' agnostic resolvent-based expansion is 2.4 faster than the ``physical'' integral-based expansion.
}

\section{Conclusion}
The quasi normal mode expansion (QNM) has shown to be a valuable tool in numerical modeling and physical understanding of nanophotonic devices. The initial works have been performed in the context of electrodynamics and optical properties of materials. We have gradually moved towards a more abstract approach where it has appeared that several decades earlier researches have already provided very general tools: the Keldy\v{s} theorem and the Lancaster work on matrix polynomials~\cite{lancaster1985theory}. 
The use of rational interpolation to take into account time dispersion is completing the setup. As a consequence, the quasi normal mode expansions can be computed independently physical nature of the considered problem, i.e. they are physically agnostic. We have presented the implementation of this approach in the eigenproblem solver SLEPc where a module is computing directly the expansion. This is avoiding a lot of heavy data transfers and storage to the physical modeling level. 
We have taken as benchmark two photonic problems computed with the FEM solver GetDP. The first one is a simple 2D academic problem and the second one is a more realistic 3D example, a periodic open structure typically encountered in electromagnetism. Compared with the implementation at the ``physical level'', the two approaches of the expansions are equivalent in accuracy but the ``agnostic approach'' is faster. 
Finally, the output of this research is a very general open source tool that may be used outside its initial application domain. With this tool, considering a new type of dispersive eigenproblem,   no extra work is required to code the expansion beyond the set up of the eigenproblem itself.

\section*{Acknowledgements}
Jose E.\ Roman was partially supported by the Spanish Agencia Estatal de Investigaci{\'o}n under grant PID2019-107379RB-I00 / AEI / 10.13039/501100011033,

\bibliographystyle{plain}
\bibliography{biblio-AN2020.bib}

\end{document}